\documentclass[conference,a4paper]{IEEEtran}
\usepackage{times,amsmath,epsfig,amsfonts,amssymb,graphicx,url,cite,subfigure,colortbl,bm,upgreek,soul,float}
\usepackage{enumerate}
\usepackage{enumitem}
\usepackage{subfigure}
\usepackage{multirow}
\usepackage{booktabs}
\usepackage{threeparttable}
\definecolor{kugray5}{RGB}{224,224,224}
\usepackage[keeplastbox]{flushend}
\usepackage{pdfpages}
\usepackage{pifont}
\usepackage[left=1.41cm,right=1.44cm,top=1.91cm,bottom=4.2cm,bindingoffset=0cm]{geometry}

\begin{document}
\title{Large-Scale Parameters of Spatio-Temporal Short-Range Indoor Backhaul Channels at 140 GHz}
\author{Sinh L. H. Nguyen$^{\dagger}$, Katsuyuki Haneda$^{\dagger}$, Jan J\"arvel\"ainen$^{\dagger}$, Aki Karttunen$^{\dagger}$ and Jyri Putkonen$^{*}$\\
\IEEEauthorblockA{$^{\dagger}$Aalto University, School of Electrical and Engineering, Finland, katsuyuki.haneda@aalto.fi\\
$^{*}$Nokia Bell Labs, Finland}}

\maketitle

\begin{abstract}
The use of above-$100$~GHz radio frequencies would be one of promising approaches to enhance the fifth-generation cellular further. Any air interface and cellular network designs require channel models, for which measured evidence of large-scale parameters such as pathloss, delay and angular spreads, is crucial. This paper provides the evidence from quasi-static spatio-temporal channel sounding campaigns at two indoor hotspot (InH) scenarios at $140$~GHz band, assuming short-range backhaul connectivity. The measured two InH sites are shopping mall and airport check-in hall. Our estimated omni-directional large-scale parameters from the measurements are found in good match with those of the Third Generation Partnership Project (3GPP) for new radios (NR) channel model in InH scenario, despite the difference of assumed link types and radio frequency range. The 3GPP NR channel model is meant for access links and said to be valid up to $100$~GHz, while our measurements cover short-range backhaul scenarios at 140 GHz. We found more deviation between our estimated large-scale parameters and those of the 3GPP NR channel model in the airport than in the shopping mall. 
\end{abstract}

\section{Introduction}
The commercial deployment of fifth-generation (5G) cellular wireless has already commenced in different parts of the world. 
Conceptualization of future generation wireless, probably called beyond-5G or sixth-generation, has already been in progress. 
One of core technological elements for future-generation wireless systems may be exploitation of unexplored frequency bands such as above-100 GHz.
It is not yet known if such high frequency bands are useful for cellular applications or not. It may be restricted to a short-range, back/front-haul and extremely high-data-rate communications such as data center applications due to physically small but electrically large antennas and high diffraction losses. Small cellular coverage for hotspots may also be possible.

Radio channel models lay foundation in feasibility study of radio systems to be designed and deployed. The one developed by Third Generation Partnership Project (3GPP) is one of the most widely used models across industry for standardization of cellular wireless. The latest channel model for New Radio (NR)~\cite{3GPP_TR38901} is said to work up to $100$~GHz, while their channel model parameters were derived from real-world measurements up to $83$~GHz carrier frequency.
Above $83$~GHz, there are handful of works reporting wave propagation and wave-material interaction. For example,
Piesiewicz {\it et al.}~\cite{Piesiewicz07_TAP} analyze reflections from materials commonly found in living environments, e.g., plasterboard and wall papers between $100$~GHz and $1$~THz, taking into account their surface roughness. 
Losses of wave penetration through materials in our living spaces are reported in~\cite{Kokkoniemi16_EuCAP,Kokkoniemi16_GSMM,Xing18_Globecom,Petrov20_EuCAP}, clearly indicating increase of the penetration loss as the frequency goes up to $10$~THz. 
Rappaport {\it et al.}~\cite{Rappaport19_Access} provide mathematical modeling of scattering from a drywall.
Abbasi {\it et al.}~\cite{Abbasi20_WCNCW} measure radar cross sections of human body at $140$ and $220$~GHz, indicating the greater cross section as the frequency increases. 

A very few short-range multipath channel measurements have also been reported for above-$83$~GHz frequencies, including those by Hanssens {\it et al.}~\cite{Hanssens18_IETMAP} analyzing specular and diffusive multipath propagation mechanisms at $94$~GHz and reporting dominance of specular mechanism and hence small delay spread up to $15$~ns.
Similar level of delay spread was reported by Pometcu and D'Errico~\cite{Pometcu19_EuCAP} for transmit-receive (Tx-Rx) distances up to $11$~m, according to their indoor measurements covering $126$-$156$~GHz frequencies. The work also shows slightly larger pathloss exponent compared to below-100 GHz frequencies at the same indoor site.
Challita {\it et al.}~\cite{Challita18_Access} study indoor spatio-temporal channels at $94$~GHz based on measurements, showing small-scale characterization such as Rician K-factors and antenna correlation.
Vitucci {\it et al.}~\cite{Vitucci18_PIMRC} compare angular profiles of indoor multipath propagation at $10$, $60$ and $300$ GHz, showing that observed power angular spectrum are consistent across the tested radio frequencies with clearer distinction of specular multipath components against diffusive multipaths at higher frequencies.
Guan {\it et al.}~\cite{Guan19_TTST} report multipath propagation channels between a train wagon and outside at $300$~GHz, observing a very limited number of multipath. Cheng and Zaji\'c~\cite{Cheng20_TAP,Cheng20_Access,Cheng19_USNC-URSI} study pathloss and delay dispersion characteristics of short range radio links at $300$~GHz assuming applications to data center scenarios. The works are based on solid measurements of up to $2.1$~m Tx-Rx separation. 
Abbasi {\it et al.}~\cite{Abbasi20_EuCAP} reports a short-range indoor channel properties for $140$-$220$~GHz, indicating dominance of line-of-sight (LOS) signal in path gain. 
The existing papers of indoor channel measurements for above-100 GHz cellular applications, e.g., indoor hotspots, are still insufficient for comprehensive analyses of such channels. We need to accumulate experimental evidence at above-100 GHz frequencies to study if the existing channel models can reproduce the observed reality of multipath propagation. 

The present paper continues analyses provided in earlier publication of the same authors~\cite{Nguyen18_EuCAP}, where large- and small-scale parameters of $28$ and $140$~GHz radio channels are compared for a shopping mall scenario. The present paper focuses on $140$~GHz measurements and discuss additional measurements from the same shopping mall and airport check-in hall, which have not been analyzed yet. Both measurements are conducted with short-range backhaul quasi-static scenarios by setting antenna heights above human heights and hence avoiding their influence on links. When we compare our estimated omni-directional large-scale parameters with the indoor channel model in 3GPP NR~\cite{3GPP_TR38901}, good agreement was observed despite differences of assumed link types and applicable radio frequency range.

The rest of the paper is organized as follows: Section~\ref{sec:sounder} details our channel sounder and measurement sites. Section~\ref{sec:results} provides analysis of large-scale parameters according to our $140$~GHz indoor channel measurements, and finally we conclude our paper in Section~\ref{sec:conclusion}. 

\begin{figure}[t]
	\begin{center}
		\includegraphics[scale=0.35]{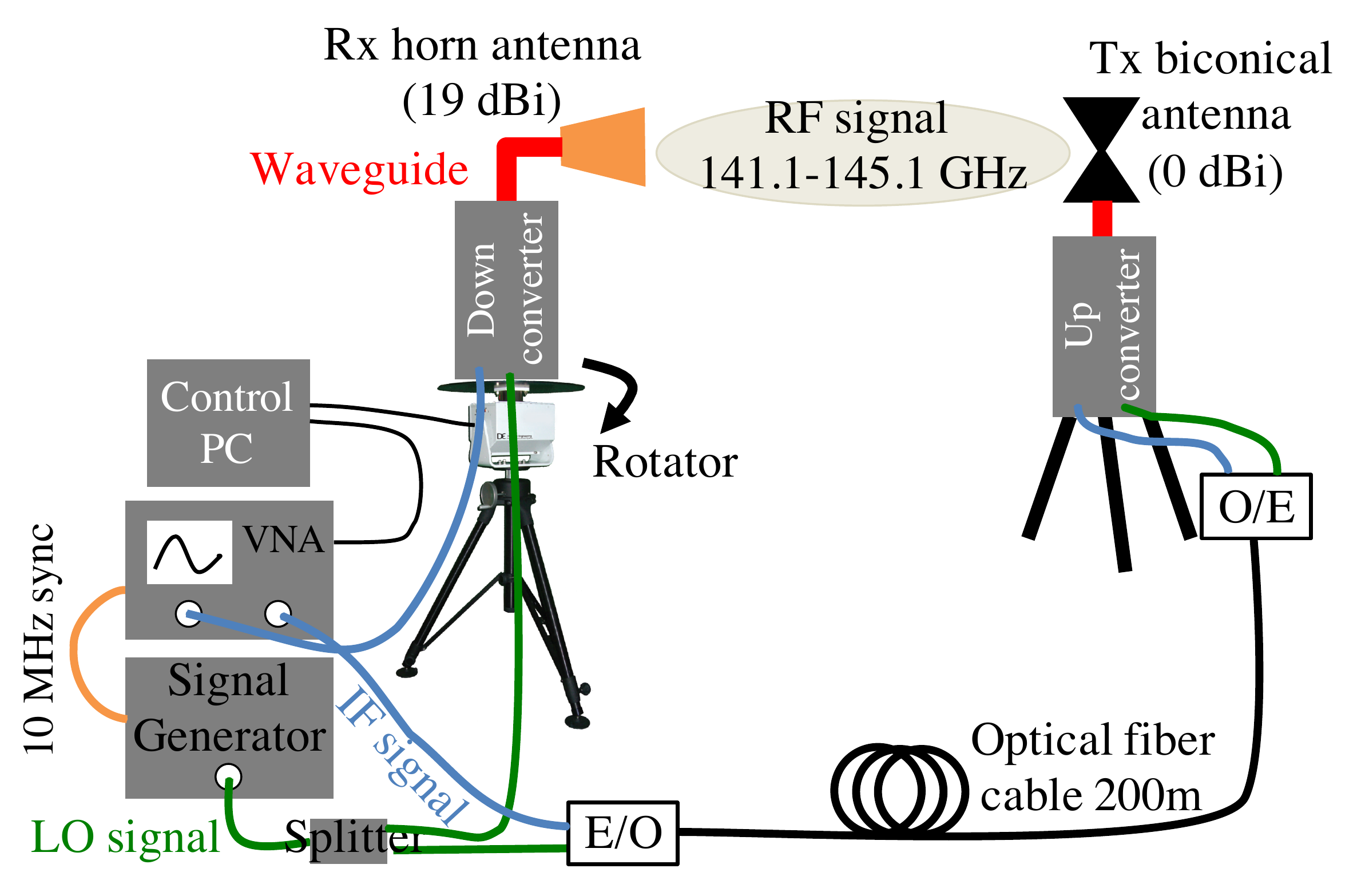}
		\caption{Schematic of channel sounder setup.}
		\label{fig:channel_sounder_140GHz}
	\end{center}
\end{figure}

\begin{table}[thbp]
	\centering
	\caption{Instruments and components consisting of the channel sounder}
	\label{tab:summary_sounder}
	\begin{tabular}{l|l}
		{\bf Instrument} & {\bf Model}\\ \hline
		VNA 			& Keysight ENA E8363A \\ \hline
		LO 				& Rohde \& Schwarz SMR 60 \\ \hline
		Rotator 		& Diamond Engineering 6100 \\ \hline
		Up-/down-converter	& Virginia Diodes MixAMC 297/MixAMC 298 \\ \hline
		RF-over-fiber & Miteq LBL-50K4P5G (IF channel) \\
					  & Miteq SCML-100M18G (LO channel)\\ \hline
		Optical fiber & Milcon P/N 2060115 \\ \hline
		Calibration attenuator & Elmika VA-02E \\ \hline
		Tx antenna & Mi-Wave OmniDirectional Antenna WR-6\\ \hline
		Rx antenna & Flann microwave custom horn antennas \\
		\end{tabular}
\end{table}

\section{Channel sounding}
\label{sec:sounder}
This section provides an overview of measurement apparatus for channel sounding, along with two indoor sites where the measurements were performed.

\subsection{Measurement apparatus and its instability}
$140$~GHz spatio-temporal channel sounder in Aalto University, Finland, uses a vector network analyzer (VNA) with up- and down-converters for frequency extension and radio frequency (RF)-over-fiber system for range extension as illustrated in Fig.~\ref{fig:channel_sounder_140GHz}~\cite{Nguyen18_EuCAP}. The sounder consists of precision instruments and RF components summarized in Table~\ref{tab:summary_sounder}. The up- and down-converters include $\times 12$ frequency multiplier for local oscillator (LO) signals that are mixed with intermediate frequency (IF) signals from the VNA. The RF-over-fiber system consists of two sets of laser diodes as electric-to-optical (E/O) converter and photo diodes for optical-to-electric (O/E) conversion. The two sets are intended for sending IF and LO signals, which are multiplexed in a single multi-mode optical fiber of $200$~m length. The RF-over-fiber system improves a dynamic range of the sounder by $30$~dB at $10$~m distance between the transmit (Tx) and receive (Rx) antennas, compared to using RF cables to feed the IF and LO signals to the up-converter. We chose an omni-directional bicone and directional horn antennas as the Tx and Rx antennas. The Rx horn antenna was rotated on the azimuth plane to obtain angularly resolved channel impulse responses.

As discussed in~\cite{Nguyen18_EuCAP}, transfer functions of the sounder is unstable over time just after instruments are turned on. The instability originates from the LO channel of RF-over-fiber system, which is then magnified at the frequency multiplier in the up- and down-converters. It turned out during laboratory tests that transfer functions of the channel sounder become stable after the instruments are on for two hours, giving uncertainty of $0.2$~dB at median level and of $2$~dB in the worst case of a peak gain of channel impulse responses, according to back-to-back calibration measurements of the sounder. The measurements were performed by connecting the RF output of the up-converter and RF input of the down-converter through an attenuator to avoid overloading the down-converter. The calibration measurements were performed right before and after field measurements, in addition to laboratory tests, in order to estimate transfer functions of the sounder.

\begin{table}[t]
	\centering
	\caption{Specifications and parameter settings in channel sounding}
	\label{tab:summary_measurements}
	\begin{tabular}{l|l}
		{\bf Properties}	& {\bf Values}\\ \hline
		IF, power		& $0.1$-$4.1$~GHz, $5$~dBm\\ \hline
		LO frequency, power	& $11.75$~GHz, $-2$~dBm  \\ \hline
		RF, power to Tx antenna & $141.1$-$145.1$ GHz, $-7$ dBm\\ \hline
		PDP dynamic range	& $125$ dB\\ \hline
		Tx / Rx height		& 1.9 / 1.9 m (shopping mall) \\
		& 1.7 m on the second floor / \\
		& 2.1 m on the third floor (airport) \\ \hline
		Link distance range	& $3-65$ m (shopping mall) \\
		& $15-51$ m (airport) \\ \hline
		Tx antenna pattern	& 0 dBi gain, $60^{\circ}$ elevation beamwidth \\ \hline
		Rx antenna pattern	& 19 dBi gain, $10^{\circ}$ azimuth beamwidth \\
		& $40^{\circ}$ elevation beamwidth
	\end{tabular}
\end{table}

\begin{figure}[t]
	\begin{center}
		\subfigure[]{\includegraphics[scale=0.35]{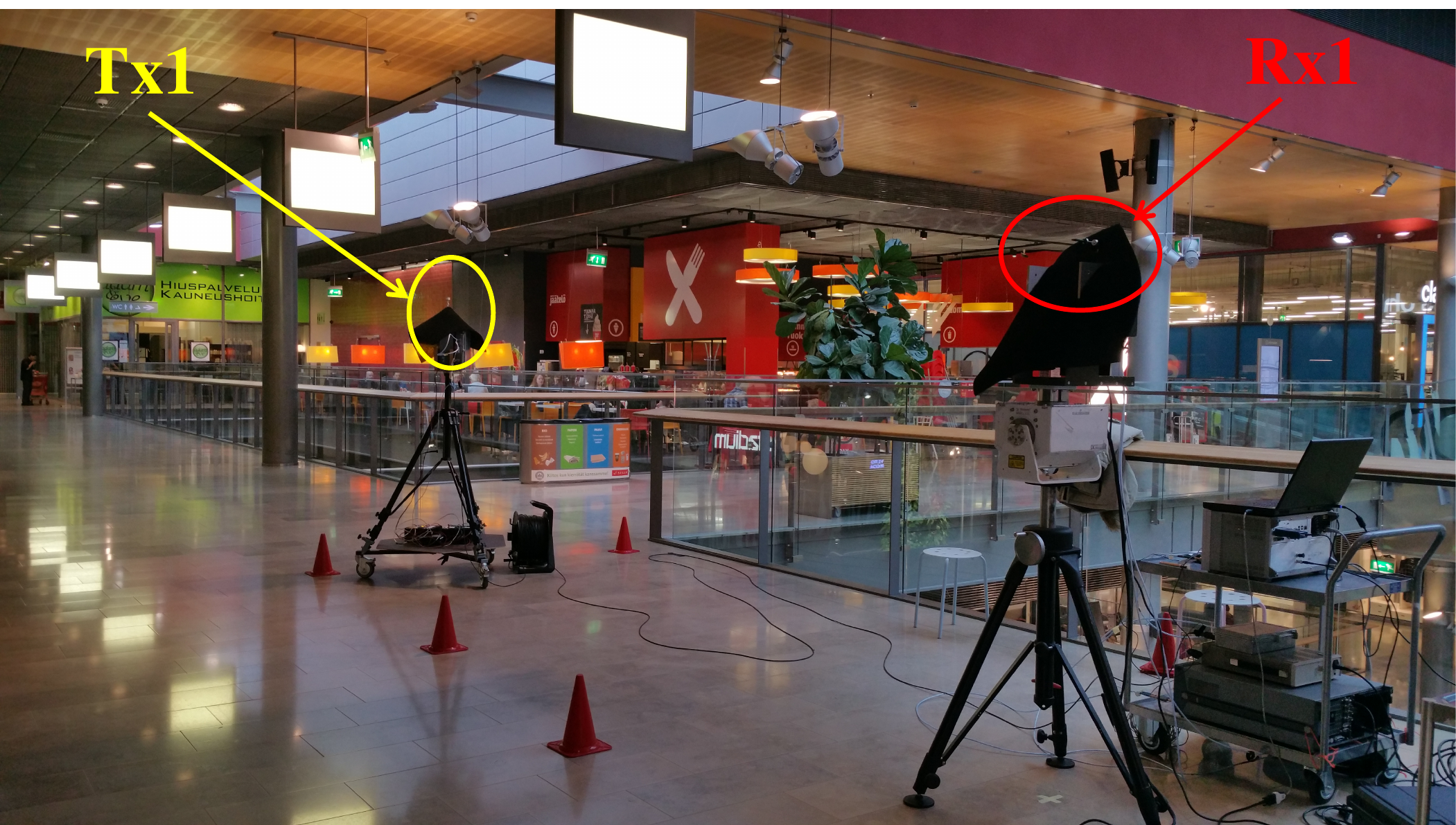}
			\label{fig:measurements_Sello_photo2}}
		\subfigure[]{\includegraphics[scale=0.4]{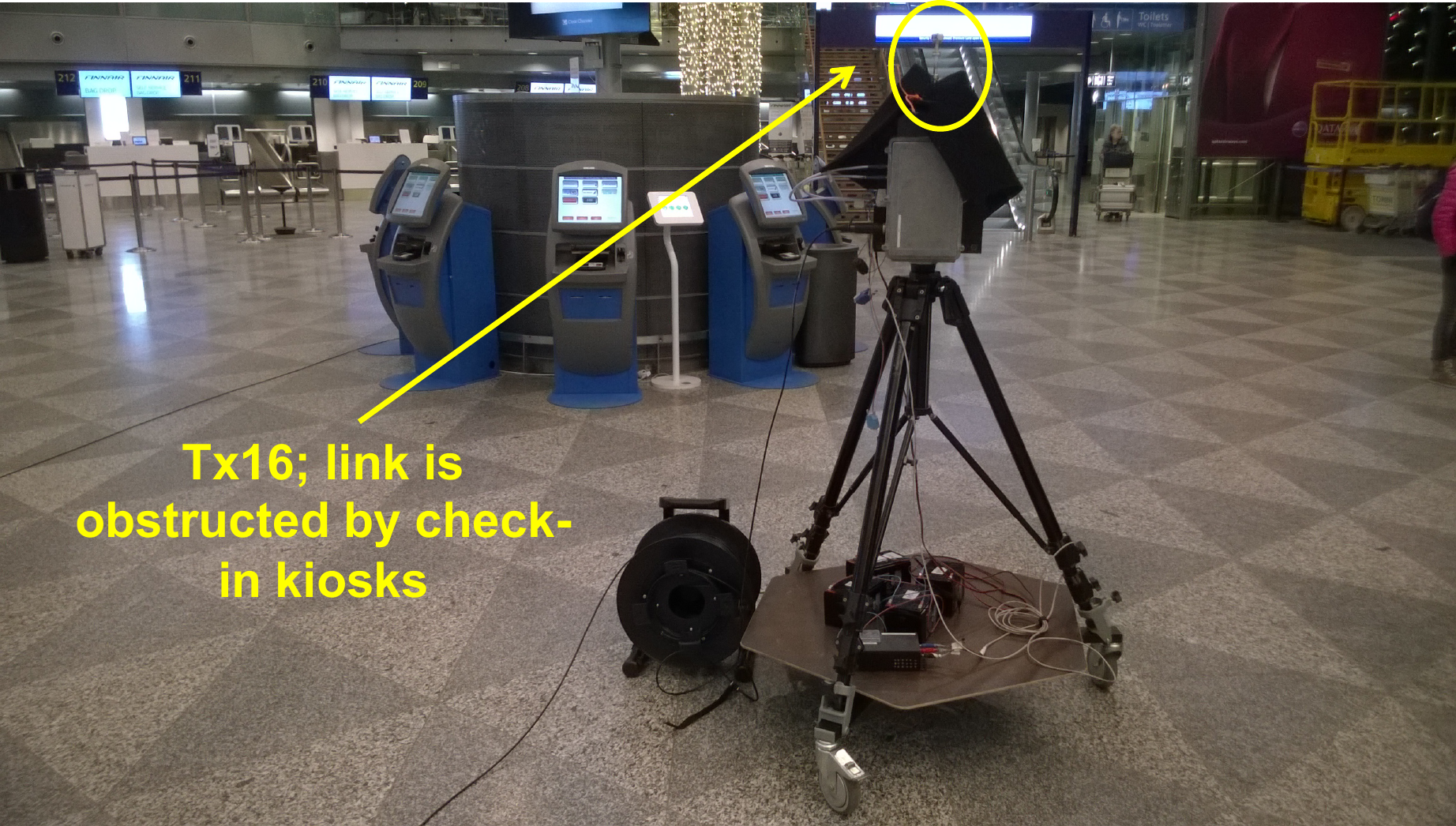}
			\label{fig:measurements_airport_photo}}
		\caption{Photos of {\it quasi-static} channel measurements without moving people interfering the link. (a) Shopping mall and (b) airport.}\label{fig:photo}
	\end{center}
\end{figure}

\begin{figure*}[t]
	\begin{center}
		\subfigure[]{\includegraphics[scale=0.48]{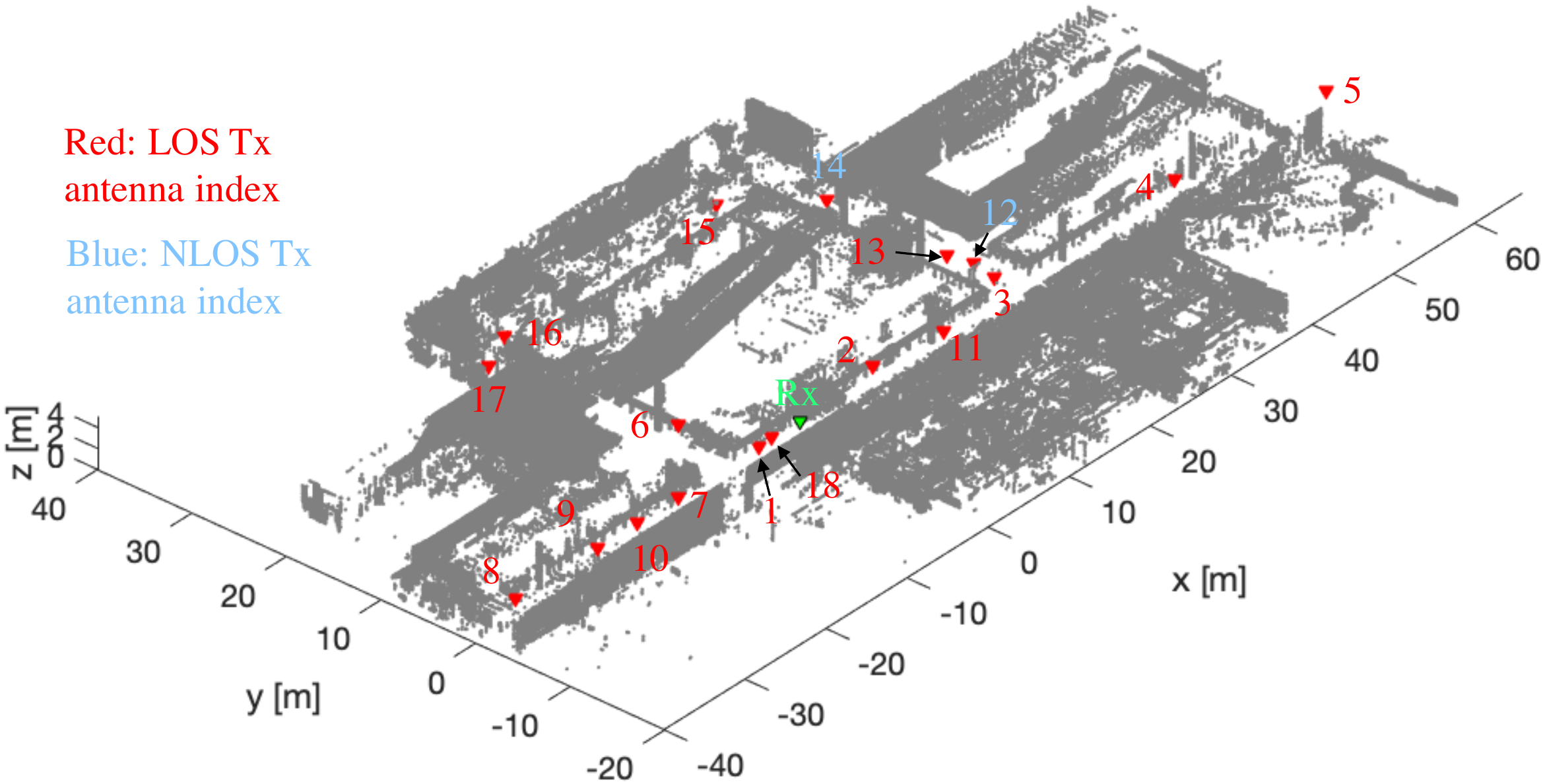}
			\label{fig:measurements_Sello_floor_plan}}
		\subfigure[]{\includegraphics[scale=0.44]{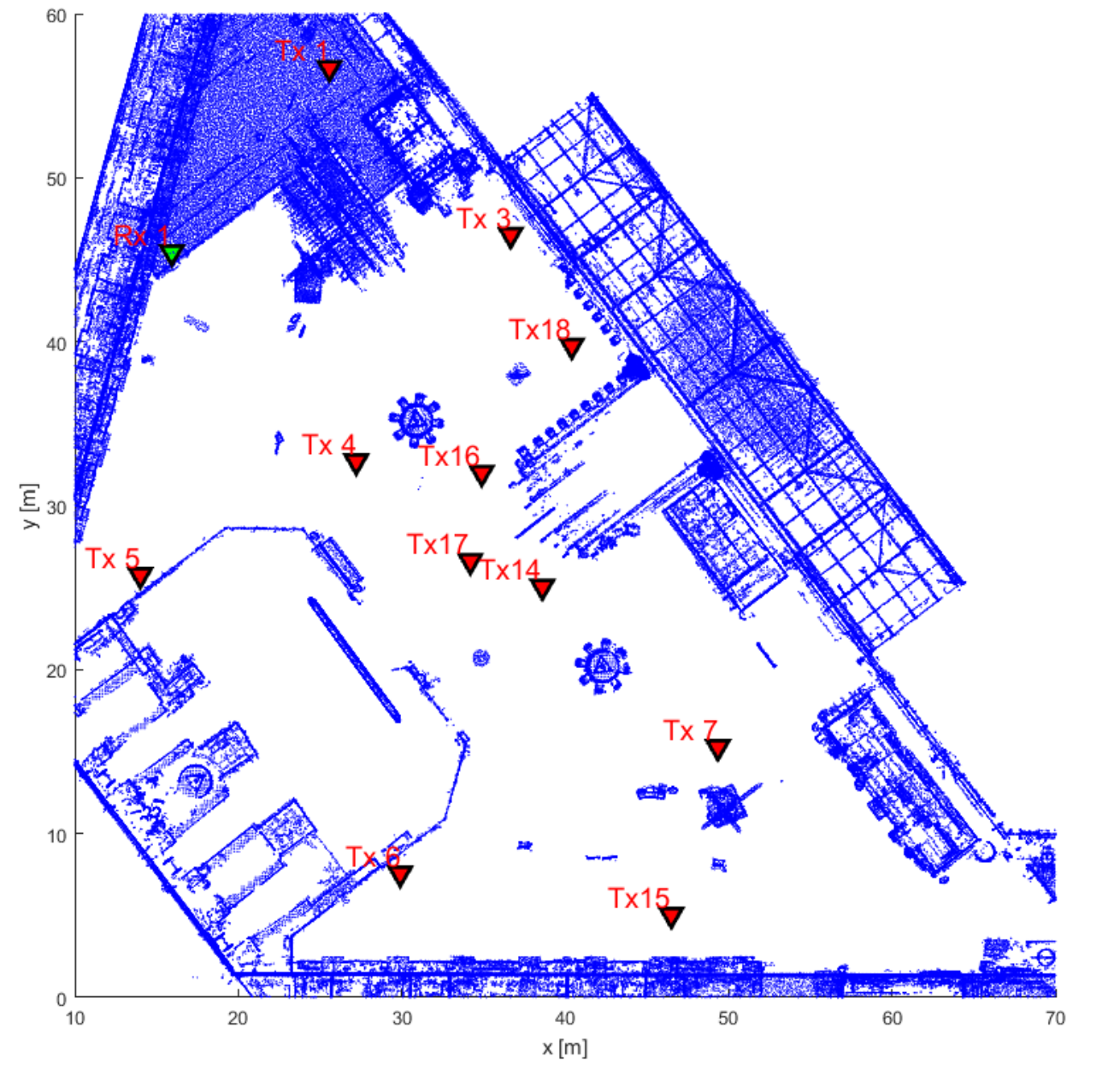}
			\label{fig:measurements_airport_floor_plan}}
		\caption{Maps of the Tx and Rx positions on (a) third floor of a shopping mall and (b) airport check-in hall for short-range backhaul channel sounding.}\label{fig:floor_plan}
	\end{center}
\end{figure*}

\subsection{Measurement sites}
\subsubsection{Shopping mall}
The first measurement site was a shopping mall ``Sello" in Espoo, Finland. It is a modern, four-story building with approximate dimensions of $120 \times 70~{\rm m}^2$ and has a large open space in the middle as seen in Fig.~\ref{fig:measurements_Sello_photo2}. The floor plan of the measurements are shown in Fig.~\ref{fig:measurements_Sello_floor_plan}. 
In total, $18$ Tx-Rx links were measured, with the Rx antenna fixed at a single location. The Tx antenna was moved along the corridor and around the open space. The antenna locations were chosen so that time-varying nature of the channels is minimized, such as link blockage due to a human body. We furthermore performed the measurements during there was no moving people or objects interfering the measured channels, leading to {\it quasi-static channels}. Both Tx and Rx antennas were elevated to $1.9$~m high above the floor, and the Tx-Rx distance ranged from $3.9$ to $65.2$~m. At two Tx antenna locations, LOS to Rx antenna was obstructed by static objects, i.e., pillar or escalator. In each Tx-Rx link measurement, the Rx horn antenna was rotated on the azimuth plane with $5^{\circ}$ step across $360^{\circ}$ to obtain angularly resolved channel impulse responses. While our earlier publication~\cite{Nguyen18_EuCAP} reports analysis of measurements performed for only $8$ links, i.e., Tx11-Tx18, the present paper is based on measurements of $18$ links, i.e., Tx1-Tx18.

\subsubsection{Airport check-in hall}
The second measurement site was a check-in hall of Helsinki Vantaa Airport, Terminal 2. The hall looks like Fig.~\ref{fig:measurements_airport_photo}, while the floor plan of the site along with Tx and Rx antenna locations are shown in Fig.~\ref{fig:measurements_airport_floor_plan}. Similarly to the shopping mall measurements, the measurements were performed during absence of moving people or objects interfering the measured channels, leading to {\it quasi-static channels}. Altogether 11 Tx antenna locations were considered across the hall, among which Tx1 was at a terrace overlooking the hall. The Rx antenna was fixed at the same terrace throughout the measurements. The terrace is $3.6$~m higher than the hall, making Tx1 and Rx antennas higher than the Tx antennas on the hall by $4.0$~m. Antenna heights are summarized in Table~\ref{tab:summary_measurements}. The Tx-Rx distance ranged from $15$ to $51$~m. All the measured links have LOS between the Tx and Rx antennas, except for Tx16 where the LOS link is obstructed by a self-check-in machine. Similarly to the shopping mall measurements, Rx horn antenna was rotated across azimuth angles with $5^{\circ}$ steps for $360^{\circ}$ range for Tx1-Rx link. In other links, the angular range is limited to $0^{\circ}$-$20^{\circ}$ and $240^{\circ}$-$360^{\circ}$ as the other azimuth angles point the horn antenna to a wall just behind the Rx location. The azimuth angles are defined according to the $x$-$y$ coordinate system of Fig.~\ref{fig:measurements_airport_floor_plan}.

\begin{figure*}[t]
	\begin{center}
		\subfigure[]{\includegraphics[scale=0.36]{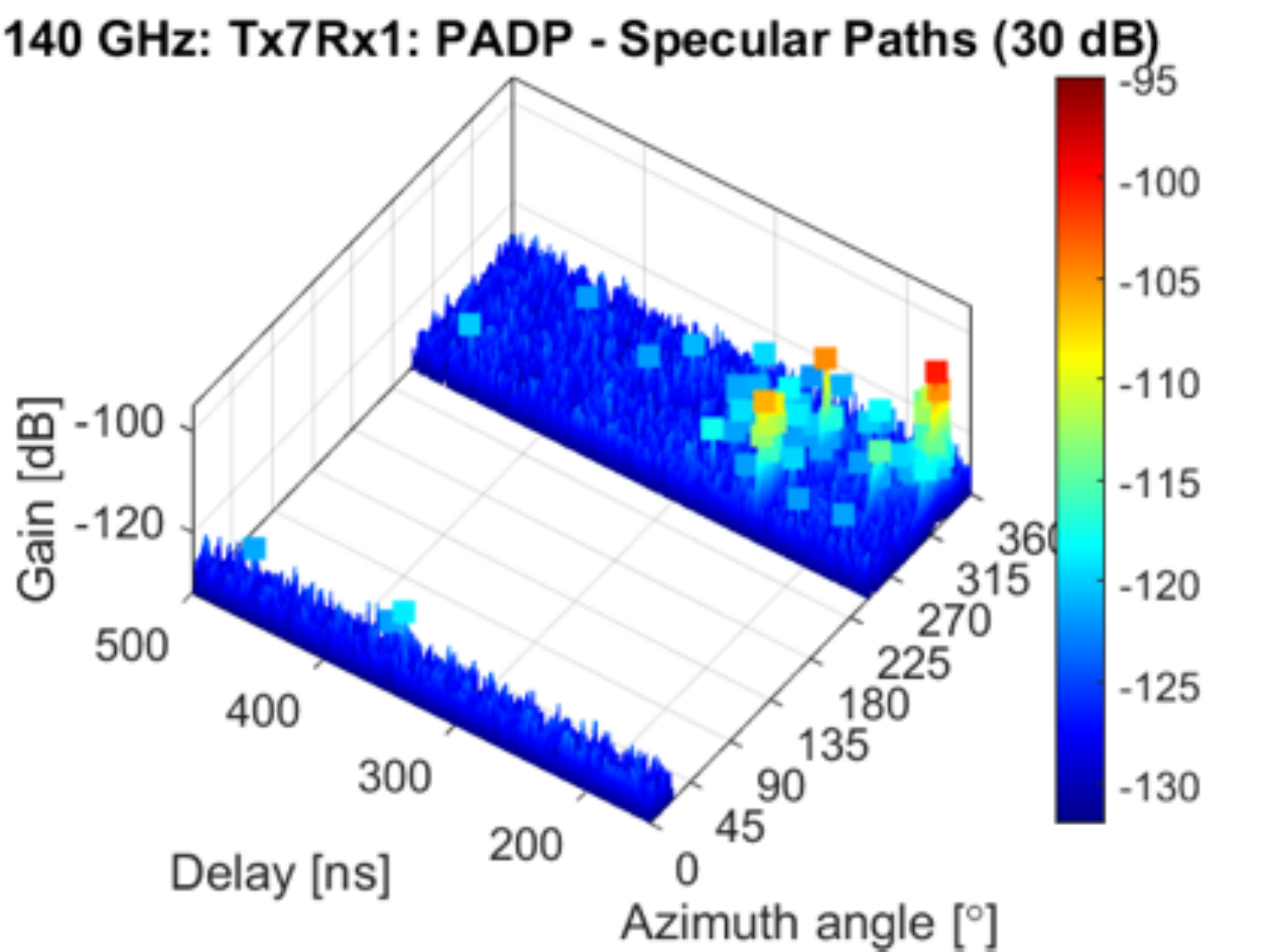}
			\label{fig:airport_Tx7_PADP}}
		\subfigure[]{\includegraphics[scale=0.5]{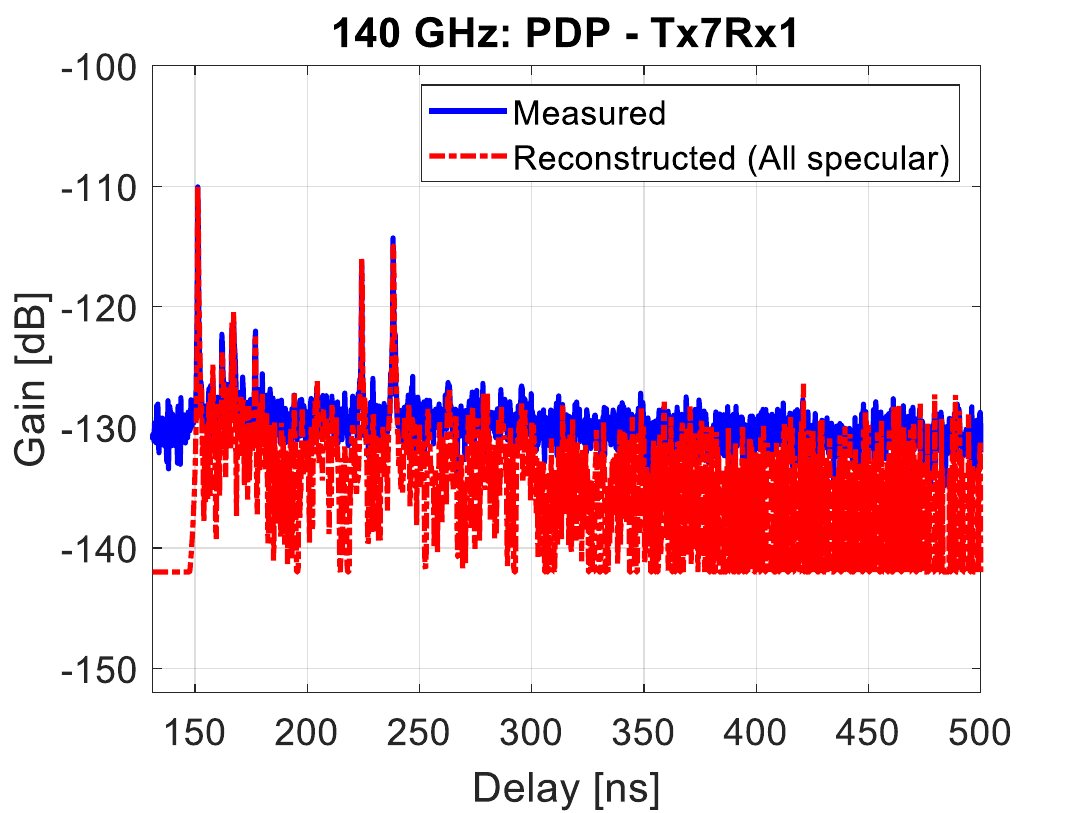}
			\label{fig:airport_Tx7_PDP}}
		\subfigure[]{\includegraphics[scale=0.5]{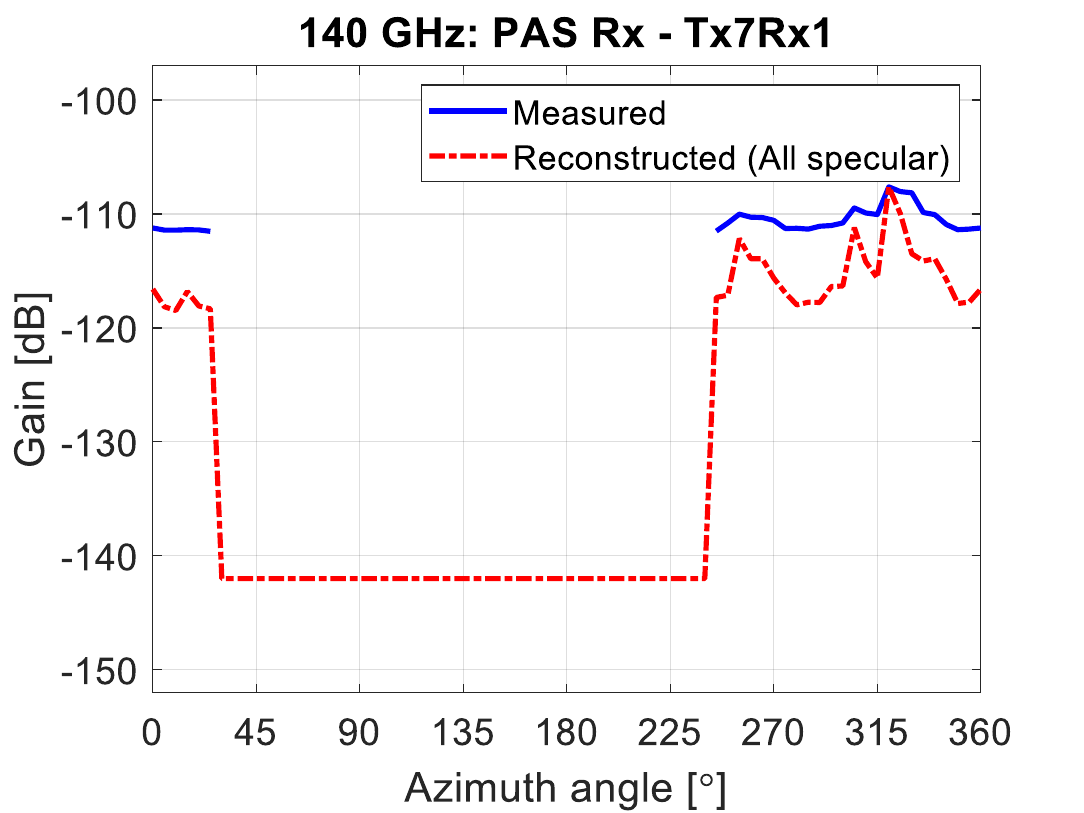}
			\label{fig:airport_Tx7_PAS}}
		\caption{Observed channel profiles in the furthest link between Tx and Rx in airport. (a) PADP, (b) PDP and (c) PAS. In (a), no data exist between $20^\circ$ and $240^\circ$ of azimuth angles.}\label{fig:airport_Tx7}
	\end{center}
\end{figure*}

\section{Characterization of multipath channels}
Three omni-directional large-scale parameters were analyzed based on our indoor multipath measurements at $140$~GHz band, i.e., pathloss, delay spread and azimuth angular spread. To this end, we first approximate the measured channels using {\it band-} and {\it antenna aperture unlimited} expression of the same, defined as
\begin{equation}
\label{eq:cir}
 h(\phi, \tau) = \sum_{l=1}^L \sqrt{G_{\rm a}} \alpha_l \delta( \phi - \phi_l) \delta( \tau-\tau_l),
\end{equation}
where $G_{\rm a}$, $\alpha$, $\phi$, $\tau$ are combined gains of the Tx and Rx antennas, path amplitude, azimuth angle of arrival and propagation delay; subscript $(\cdot)_l$ denotes an $l$-th multipath, $1 \le l \le L$. There is an apparent difference between~\eqref{eq:cir} and measured channels, exemplified in Fig.~\ref{fig:airport_Tx7_PADP} for Tx7-Rx link from the airport measurement. What we can measure and observe is always {\it band-} and {\it antenna aperture limited} form that we call a power angular delay profile (PADP) hereinafter. The approximation of band- and antenna aperture limited measurements by the mathematically versatile band- and antenna aperture unlimited formula is performed by finding local maxima of the measured PADP, and assuming that the maxima correspond to physical multipaths, i.e., $L$ multipaths. Our earlier publication~\cite{Nguyen18_EuCAP} elaborates a simple but robust method to estimate multipath parameters, i.e., $\alpha_l$, $\phi_l$, $\tau_l$, for $1 \le l \le L$ through peak search of the PADP. The combined gains of the Tx and Rx antennas, $G_{\rm a} = 19$~dBi, are known for {\it antenna gain de-embedding} of estimated multipaths assuming that multipaths are detected at the broadside of the Tx and Rx antennas. 

The large-scale parameters of our interests are derived from the estimated multipath parameters. We use an ordinal approach~\cite{Molisch11_book} with 1) only analyzing multipaths with high-enough gain, i.e., between $\max_l |\alpha_l|^2_{\rm dB}$ and $\max_l |\alpha_l|^2_{\rm dB} - 30$~dB and 2) ensuring circular continuity for the azimuth angles when calculating the azimuth spread. However, concerning 1), since some links typically with a longer Tx-Rx distance than $30$~m have a smaller dynamic range of PADP than $30$~dB, we use the observed dynamic range of the measurements in the large-scale parameter calculation if it is less than $30$~dB. Here the dynamic range is defined by a difference between the strongest signal and noise floor levels. 

\section{Results and Discussions}
\label{sec:results}
This section first summarizes observations from our measurements in terms of PADP, PDP and PAS, along with our estimates of omni-directional large-scale parameters, i.e., pathloss, delay and azimuth angular spreads, after antenna gain de-embedding. We finally compare the obtained large-scale parameter estimates from those of a reference channel model, i.e. 3GPP NR InH model.

\subsection{Observations of multipath channels}
For illustration of measured $140$~GHz indoor multipath channels, Figs.~\ref{fig:airport_Tx7_PDP} and~\ref{fig:airport_Tx7_PAS} show power delay profiles (PDP) and power angular spectrum (PAS). They are obtained by marginal integration of the PADP, i.e., Fig.~\ref{fig:airport_Tx7_PADP}, over azimuth and delay domains, respectively. As noise power is integrated as well as multipath signal powers, the noise floor of PADP, PDP and PAS are different as apparent from Fig.~\ref{fig:airport_Tx7}. The highest and lowest noise floors of $-110$ and $-132$~dB are observed in the PAS and PADP, among the three plots.

Figure~\ref{fig:airport_Tx7} shows PADP, PDP and PAS of Tx7-Rx link at airport, with the $45.3$~m link separation. The PADP is overlaid by detected multipaths as significant local maxima, colored according to its gain. The plot shows more paths than LOS, indicating possibilities to deliver energy from one link end to another through multiple physical paths for spatial multiplexing and when LOS is blocked by, e.g., a human body. There are particularly strong paths at $250$ and $275$~ns even though they have much longer delays than other multipaths, clearly because of specular reflections from large and flat and metallic side walls of the airport. The dynamic range of the PADP is $26$~dB. Figures~\ref{fig:airport_Tx7_PDP} and~\ref{fig:airport_Tx7_PAS} include reproduced profile and spectrum according to our multipath estimates indicated as red lines. They are band- and antenna aperture-limited form derived by convolving a sinc function and antenna field pattern with~\eqref{eq:cir}. The red curves show successful detection of many distinct local maxima of the measured PDP and PAS, while they show more deviation from the measurement at low levels since they represent mainly noise. The measured curves have high noise level as a result of accumulating noise power in deriving the PDP and PAS from the PADP, while the reproduced curves are free from noise. The overall agreement between measured and reproduced curves for multipath signals prove efficacy of the multipath estimation method.

\begin{figure*}[t]
	\begin{center}
		\subfigure[]{\includegraphics[scale=0.33]{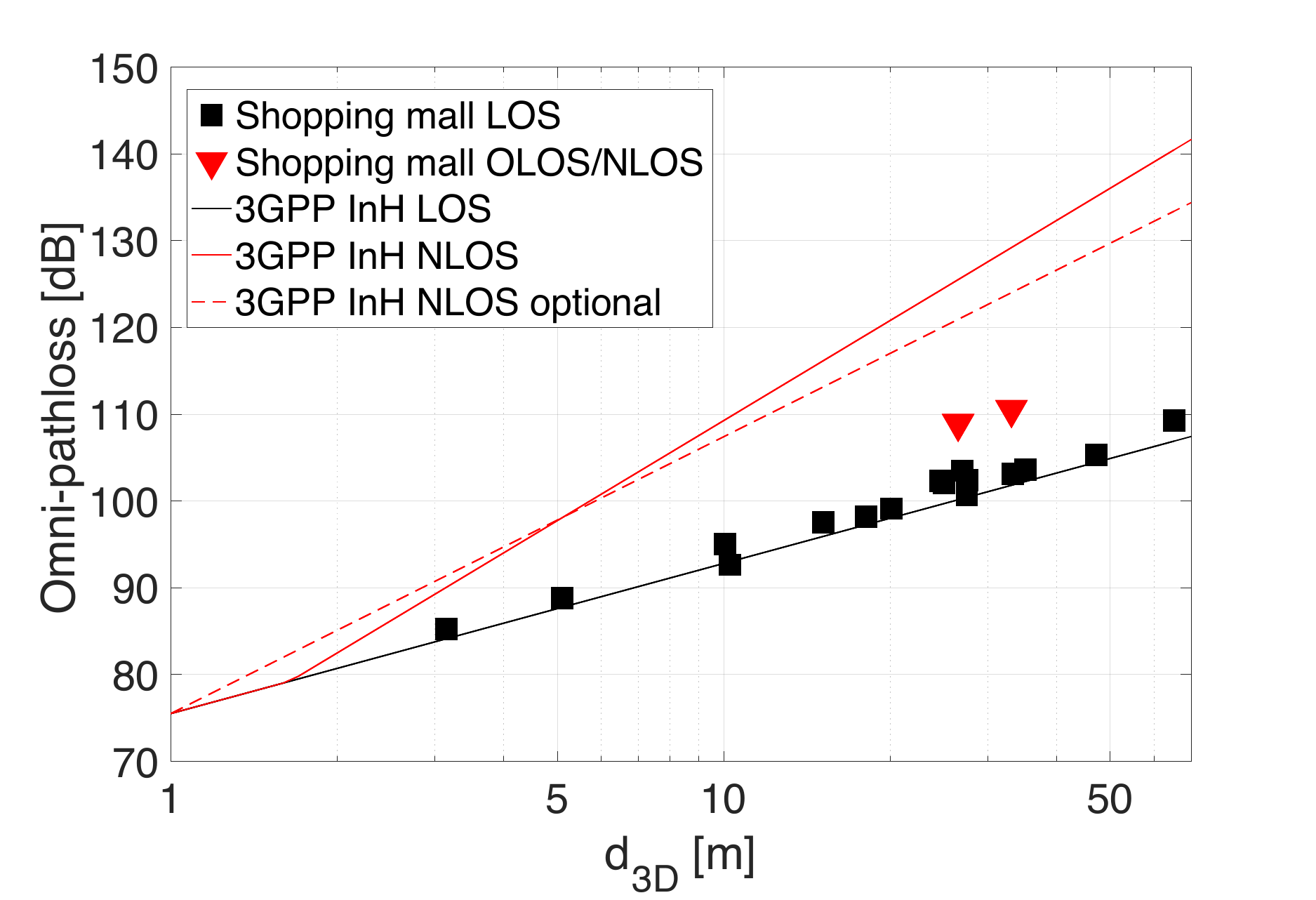}
			\label{fig:pathloss_Sello}}
		\subfigure[]{\includegraphics[scale=0.32]{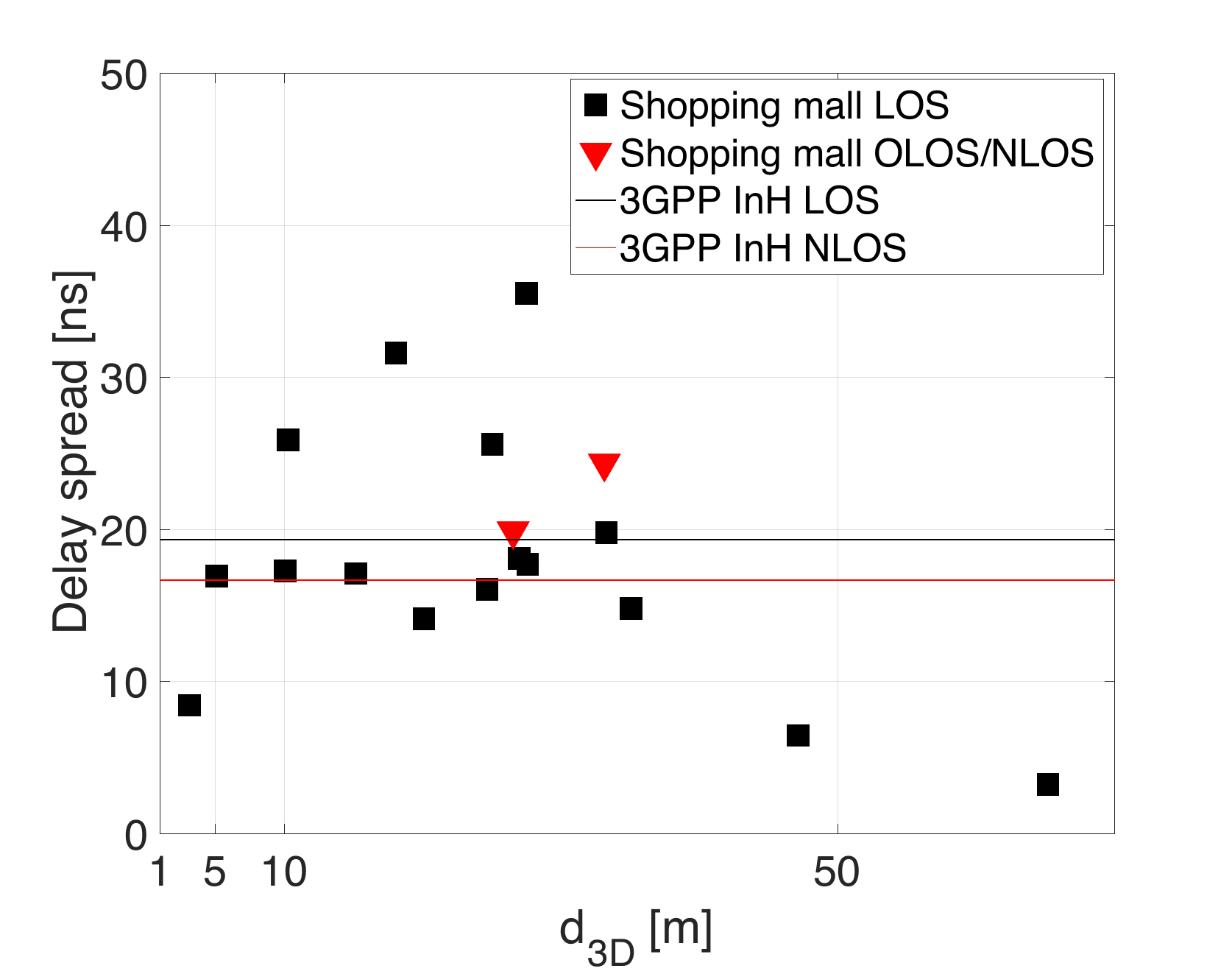}
			\label{fig:delay_spread_Sello}}
		\subfigure[]{\includegraphics[scale=0.33]{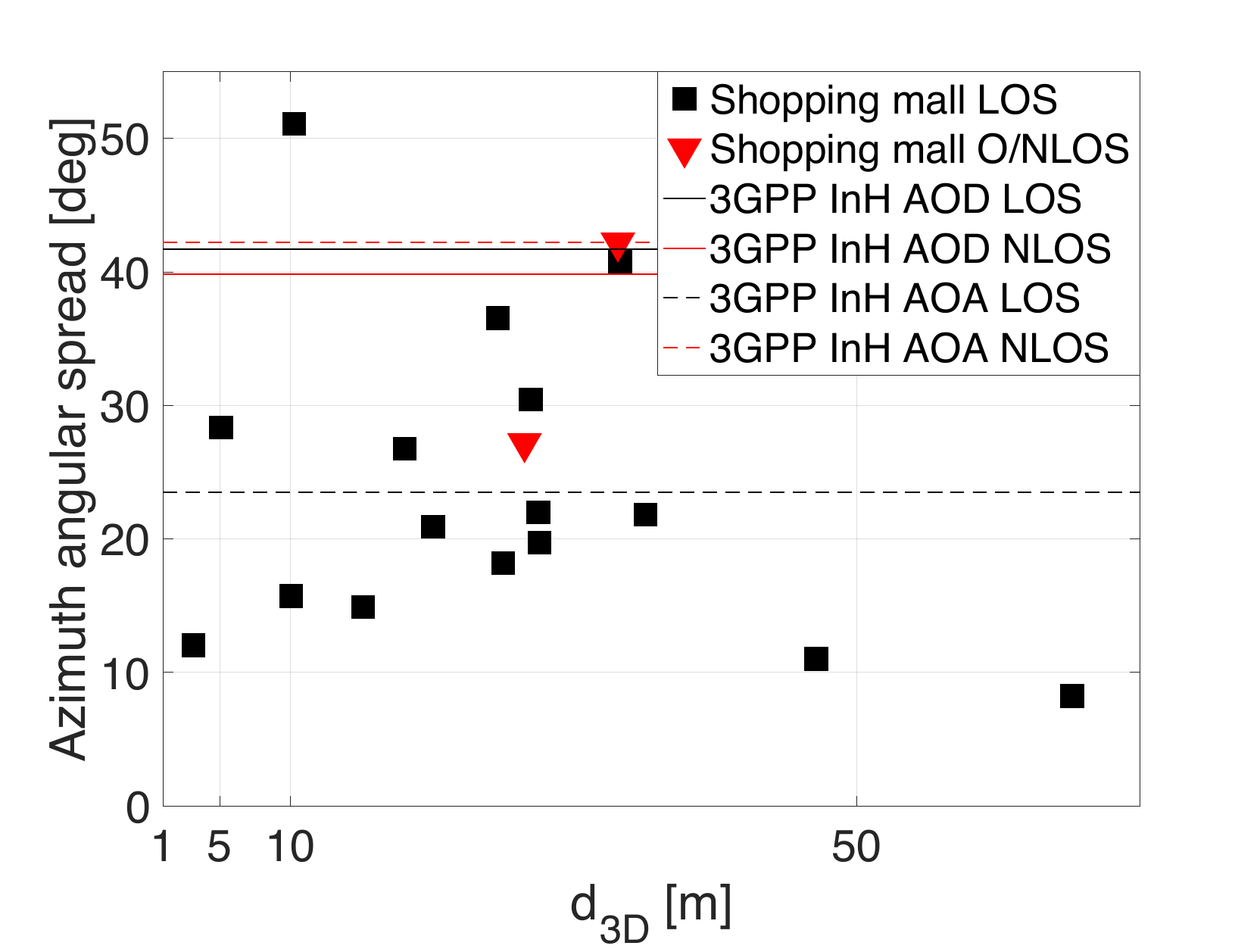}
			\label{fig:azimuth_spread_Sello}}
		\subfigure[]{\includegraphics[scale=0.34]{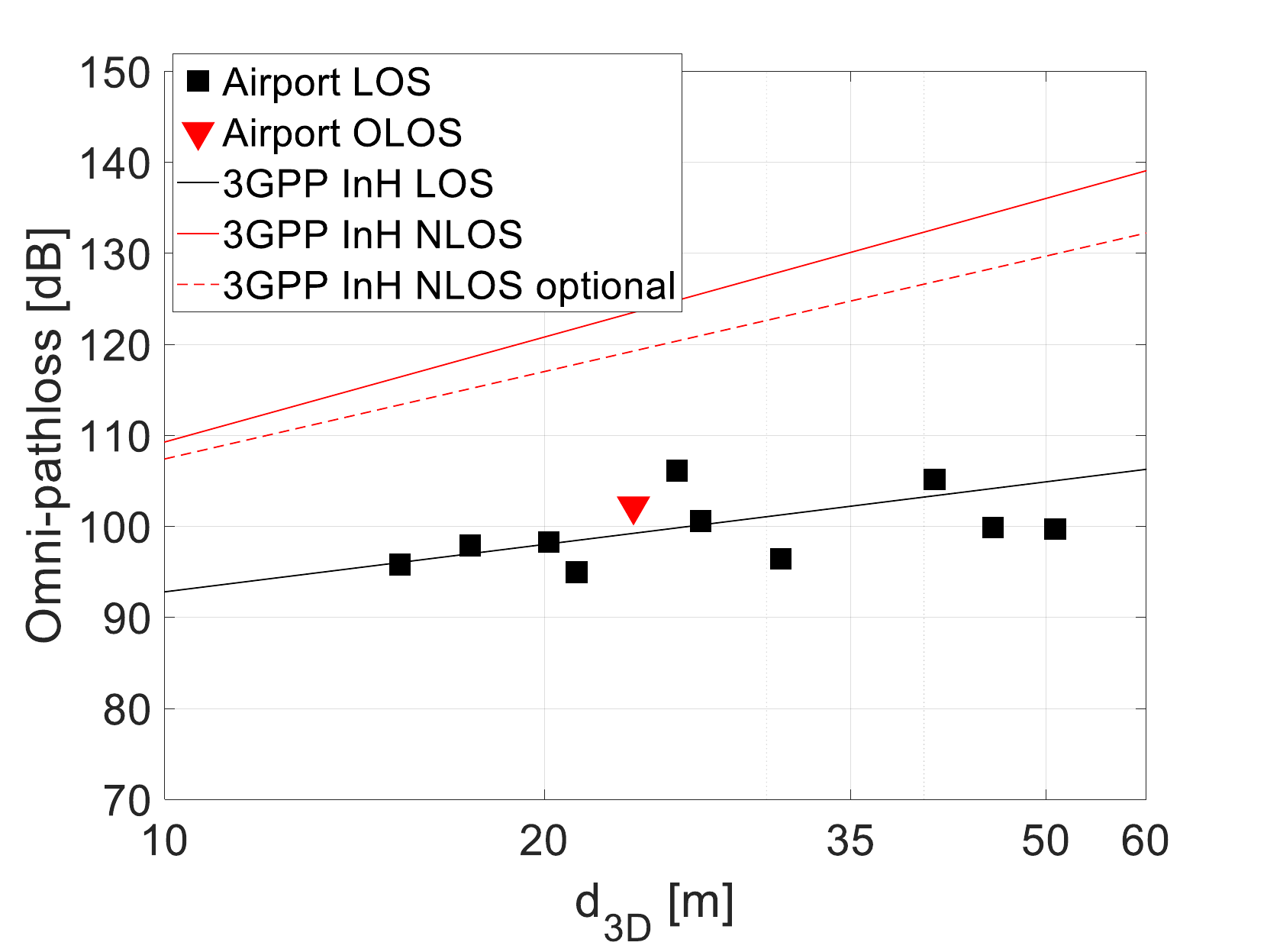}
			\label{fig:pathloss_airport}}
		\subfigure[]{\includegraphics[scale=0.38]{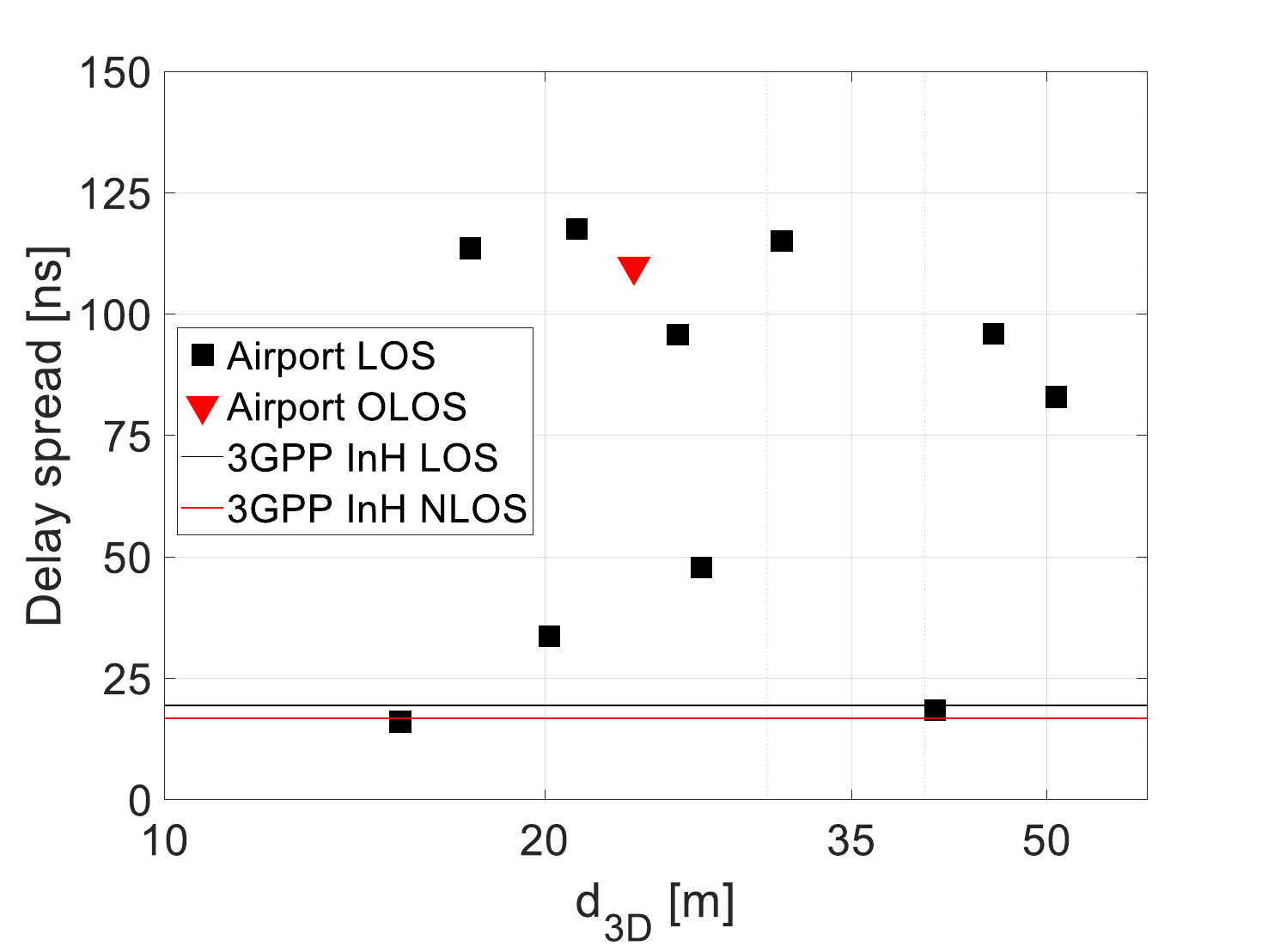}
			\label{fig:delay_spread_airport}}
		\subfigure[]{\includegraphics[scale=0.34]{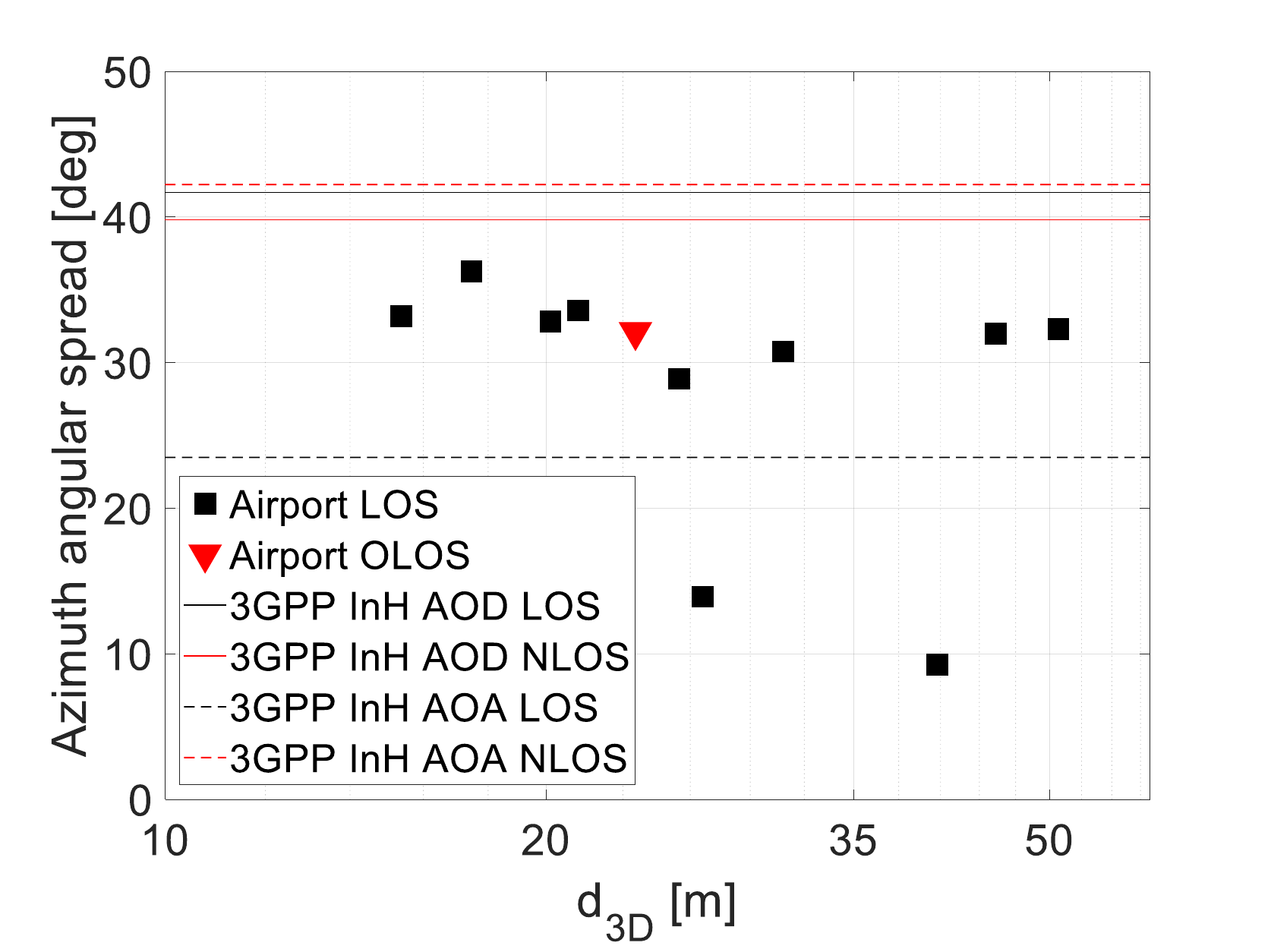}
			\label{fig:azimuth_spread_airport}}
		\caption{Large-scale omni-directional channel parameters at 140 GHz. (a) pathloss (b) delay spread and (c) azimuth spread at Rx in shopping mall and (d) pathloss (e) delay spread and (f) azimuth spread at Rx in airport.}\label{fig:large_scale}
	\end{center}
\end{figure*}

\subsection{Large-Scale Parameters}
The estimated large-scale parameters shown in Fig.~\ref{fig:large_scale}. Each plot in the figure is overlaid by reference models of 3GPP NR for the indoor hotspot (InH) scenario~\cite{3GPP_TR38901}. The model shows that all the studied large-scale parameters in this paper have frequency dependency. Though the model is verified only up to $83$~GHz measurements, we substitute $140$~GHz to the frequency term. It must also be noted that the 3GPP NR models are for access links, while our measurements assume short-range backhaul links. The major differences between the two types of links are antenna heights, where the former may have different antenna heights at link ends, while the latter may have the same elevated antenna heights than human, as performed in our channel sounding.

\subsubsection{Shopping mall}
For shopping mall scenario, the measured large-scale parameter estimates at $140$~GHz follow the 3GPP InH LOS model parameters well. The mean differences between the measurements and LOS models are $+0.9$~dB in pathloss, $+1.3$~ns in delay spread and $+18.0^\circ$ and $-0.2^\circ$ in azimuth angle-of-departure (AoD) and angle-of-arrival (AoA) spreads. The sign ``$+$" indicates that the measured values are greater than those in the model. Standard deviations are $3.6$~dB in pathloss, $8.5$~ns in delay spread and $11.6^\circ$ in azimuth spread. The largest differences of delay and angular spreads between measurements and models are at the two longest Tx-Rx links, indicating possible influence of a limited dynamic range.

\subsubsection{Airport}
For airport check-in hall, the measured pathloss follows the 3GPP NR InH LOS model parameters well. The mean difference between the measurements and LOS model is $-0.9$~dB, with standard deviation of $3.6$~dB. However, delay spread of the LOS model is too small to represent our measurements at $140$~GHz, showing a mean difference and standard deviation of $54$ and $40$~ns because of long-delayed multipaths exemplified in Fig.~\ref{fig:airport_Tx7_PDP}. Finally, the measured azimuth spread is always smaller than the LOS models probably because only a limited angular range is scanned at the wall-side Rx. Two links Tx17-Rx and Tx6-Rx show significantly smaller values than the LOS models because most multipaths arrive from a limited angular range. It must be noted again that azimuth angles of multipaths are estimated at the Rx on the terrace, which can resemble a base station overlooking the hall. The mean differences between measurements and LOS models are $-13.4^\circ$ and $4.8^\circ$ for AoD and AoA, while the standard deviation is $9.1^\circ$. The agreement between measurements and 3GPP InH LOS models is generally worse than shopping mall scenario because the airport hall may be bigger than ordinal InH scenario. More long-delayed multipaths provides greater delay spread, while longer LOS link results in the reduced angular range of multipaths. 

\section{Concluding Remarks}
\label{sec:conclusion}
Large-scale parameters of $140$~GHz InH links were reported based on spatio-temporal channel sounding at two sites, i.e., shopping mall and airport check-in hall, which covered $18$ and $11$ measured links respectively. The measurements assumed short-range backhaul scenarios and were performed during the absence of moving people so that quasi-static channel conditions were ensured for the use our channel sounder. Multipath parameters of the channels, i.e., path magnitude, AoA and propagation delay, were estimated through peak detection of the PADP. Tx and Rx antenna gains were de-embedded from the path magnitude estimates. Our large-scale parameter estimates at $140$~GHz, i.e., pathloss, delay and azimuth angular spreads, were finally compared with a reference channel model, i.e., the 3GPP NR InH LOS channel model in this paper, which was derived from measurements with RF up to $83$~GHz and nominally valid up to $100$~GHz. The comparison showed good agreement in the shopping mall, despite that the 3GPP NR InH channel model is for access links while our measurements resemble short-range backhaul links. The same comparison for airport showed more deviation between our measurements and the 3GPP NR model. The present evidence from channel sounding suggests further comparison between measured reality of multipath channels and reference channel models, strengthened by more channel sounding campaigns. Additionally, small-scale multipath characteristics, e.g., clusters, will have to be addressed at $140$~GHz to see if existing reference channel models can represent them well. Finally, validity of other possible reference channel models will also be tested against our measurements, e.g., ray-based site-specific channel models.

\section*{Acknowledgement}
The authors would like to thank Mr.\ Usman Virk, Dr.\ Mamadou Balde and Mr.\ Bi\c{c}er Sena for their help in channel sounding and laboratory tests. K. Haneda acknowledges financial support of European Union Horizon 2020 project Artificial Intelligence Aided D-band Network for 5G Long Term Evolution (ARIADNE), proposal \#871464.

\bibliographystyle{IEEEtran}
\bibliography{ref}
\end{document}